\documentclass[%
reprint,
superscriptaddress,
amsmath,amssymb,
aps,
prc,
]{revtex4-2}

\usepackage[detect-all]{siunitx}
\DeclareSIUnit{\weisskopfunit}{W.\hspace{1pt}u.}
\newcommand{\asymUnc}[4]{\ensuremath{#1\;^{+\;#2}_{-\;#3}\;\hspace{-0pt}\si{#4}}}

\usepackage[version=4]{mhchem}
\usepackage{graphicx}
\usepackage{dcolumn}
\usepackage{bm}
\usepackage{hyperref}
\usepackage{multirow}
\usepackage{tabularx}
\usepackage{float}

\usepackage{xcolor}

\newcommand{\bdown}{$B(E2)$$\downarrow$ }

\begin{document}
	
	\preprint{APS/123-QED}
	
    \title{Lifetimes of the \ce{{2}^{+}_{1}} and \ce{{4}^{+}_{1}} states of the neutron-rich nuclide \ce{^{200}Pt}}

	
	\author{C. M. Nickel}
	\email[Corresponding author: ]{cnickel@ikp.tu-darmstadt.de}
	\affiliation{
		Institute for Nuclear Physics, Dept. of Physics, Technische Universität Darmstadt, Schlossgartenstraße 9, D-64289 Darmstadt, Germany
	}
    \author{V. Werner}
    \affiliation{
		Institute for Nuclear Physics, Dept. of Physics, Technische Universität Darmstadt, Schlossgartenstraße 9, D-64289 Darmstadt, Germany
	}
    \affiliation{Helmholtz Forschungsakademie Hessen für FAIR (HFHF), Campus Darmstadt, Schlossgartenstraße 2, D-64289 Darmstadt, Germany}
    
	\author{P. R. John}
	\affiliation{
		Institute for Nuclear Physics, Dept. of Physics, Technische Universität Darmstadt, Schlossgartenstraße 9, D-64289 Darmstadt, Germany
	}
    \author{U. Ahmed}
    \affiliation{
		Institute for Nuclear Physics, Dept. of Physics, Technische Universität Darmstadt, Schlossgartenstraße 9, D-64289 Darmstadt, Germany
	}
    \affiliation{Helmholtz Forschungsakademie Hessen für FAIR (HFHF), Campus Darmstadt, Schlossgartenstraße 2, D-64289 Darmstadt, Germany}
    \author{T. Beck}
    \affiliation{
		Institute for Nuclear Physics, Dept. of Physics, Technische Universität Darmstadt, Schlossgartenstraße 9, D-64289 Darmstadt, Germany
	}
    \author{M. Boromiza}
    \author{\\C. Clisu-Stan}
    \author{A. Coman}
	\author{C. Costache}
    \author{N. M. Florea}
    \affiliation{
		Dept. of Nuclear Physics, Horia Hulubei National Institute for R\&D in Physics and Nuclear Engineering (IFIN-HH), 30 Reactorului St., RO-077125 Bucharest-M\u{a}gurele, Romania
	}    
    \author{K. E. Ide}
    \affiliation{
		Institute for Nuclear Physics, Dept. of Physics, Technische Universität Darmstadt, Schlossgartenstraße 9, D-64289 Darmstadt, Germany
	}
    \author{A. Ionescu}
    \author{R. Lic\u{a}}
    \author{N. M. M\u{a}rginean}
    \author{R. M\u{a}rginean}
    \author{A. Mitu}
    \affiliation{
		Dept. of Nuclear Physics, Horia Hulubei National Institute for R\&D in Physics and Nuclear Engineering (IFIN-HH), 30 Reactorului St., RO-077125 Bucharest-M\u{a}gurele, Romania
	}
	\author{H. Mayr}
    \affiliation{
		Institute for Nuclear Physics, Dept. of Physics, Technische Universität Darmstadt, Schlossgartenstraße 9, D-64289 Darmstadt, Germany
	}
    \author{C. Mihai}
	\affiliation{
		Dept. of Nuclear Physics, Horia Hulubei National Institute for R\&D in Physics and Nuclear Engineering (IFIN-HH), 30 Reactorului St., RO-077125 Bucharest-M\u{a}gurele, Romania
	}
    \author{R. E. Mihai}
	\affiliation{
		Dept. of Nuclear Physics, Horia Hulubei National Institute for R\&D in Physics and Nuclear Engineering (IFIN-HH), 30 Reactorului St., RO-077125 Bucharest-M\u{a}gurele, Romania
	}
	\affiliation{Czech Technical University in Prague, Institute of Experimental and Applied Physics, Husova 240/5, CZ-110\,00 Prague, Czech Republic}
    \author{S. Pascu}
	\affiliation{
		Dept. of Nuclear Physics, Horia Hulubei National Institute for R\&D in Physics and Nuclear Engineering (IFIN-HH), 30 Reactorului St., RO-077125 Bucharest-M\u{a}gurele, Romania
	}
	\author{N. Pietralla}
    \affiliation{
		Institute for Nuclear Physics, Dept. of Physics, Technische Universität Darmstadt, Schlossgartenstraße 9, D-64289 Darmstadt, Germany
	}
    \author{L. Stan}
    \affiliation{
		Dept. of Nuclear Physics, Horia Hulubei National Institute for R\&D in Physics and Nuclear Engineering (IFIN-HH), 30 Reactorului St., RO-077125 Bucharest-M\u{a}gurele, Romania
	}
	\author{T. Stetz}
    \affiliation{
		Institute for Nuclear Physics, Dept. of Physics, Technische Universität Darmstadt, Schlossgartenstraße 9, D-64289 Darmstadt, Germany
	}
    \author{A. E. Turturic\u{a}}
    \author{S. Ujeniuc}
    \affiliation{
		Dept. of Nuclear Physics, Horia Hulubei National Institute for R\&D in Physics and Nuclear Engineering (IFIN-HH), 30 Reactorului St., RO-077125 Bucharest-M\u{a}gurele, Romania
	}
	\author{A. Weber}
	\author{R. Zidarova}
	\affiliation{
		Institute for Nuclear Physics, Dept. of Physics, Technische Universität Darmstadt, Schlossgartenstraße 9, D-64289 Darmstadt, Germany
	}
	
	\date{\today}
	
	\begin{abstract}
		The lifetimes of the $2^+_1$ and $4^+_1$ states of \ce{^{200}Pt} were measured applying the recoil-distance Doppler-shift method. Excited states were populated in the \ce{^{198}Pt}(\ce{^{18}O}, \ce{^{16}O})\ce{^{200}Pt} two-neutron transfer reaction at the \SI{9}{\mega\volt} tandem accelerator at the IFIN-HH in M\u{a}gurele, Romania. The resulting $B(E2)$ values of the $2^+_1 \rightarrow 0^+_1$ and $4^+_1 \rightarrow 2^+_1$ transitions as well as the $B_{4/2}$ ratio of $\num{2.08 \pm 0.32}$ indicate the nuclear structure evolving towards sphericity when approaching the neutron shell closure at $N = 126$. The $B(E2;2^+_1 \rightarrow 0^+_1)$ values of Pt and Hg are compared to values of Te, Xe and Ba as both regions of the nuclear chart show similar structural effects.
        
	\end{abstract}
	
	\maketitle
	
	
	\section{Introduction}

    The region of the nuclear chart below \ce{^{208}Pb}, i.e., neutron-rich isotopes from Hf to Pt, has been discussed in the context of a transition from prolate to oblate shapes \cite{Jolie02,Jolie01,Stevenson01,Sarriguren01,Alkhomashi01,Nomura01}. In particular, such a transition would cross through the limit of soft triaxiality ($\gamma$-softness). While \ce{^{190}W} has recently been discussed as an isotope closest to the prolate-oblate transition \cite{Sahin24}, in particular \ce{^{196}Pt} has long been a textbook example of a $\gamma$-soft structure~\cite{Cizewski01,Casten04}, in close proximity to the O(6) dynamical symmetry limit of the interacting boson model \cite{IachelloIBM}. However, moving to heavier isotopes, the proximity of the $N = 126$ neutron shell closure takes effect, as the size of the available valence space, hence, the degree of collectivity, drops quickly towards the shell closure, and sphericity is expected to establish near $N = 126$.

    A well-known measure for the degree of collectivity is the $E2$ transition strength, in particular, of even-even nuclei. Since the reduced transition probability $B(E2)$$\downarrow~\equiv~B(E2;2^+_1 \rightarrow 0^+_1)$ exhausts most of the $E2$~transition strength, it is an indicator of the evolution of structure over a series of isotopes or, more generally, in a given region of the nuclear chart. To investigate the evolution towards the $N = 126$ shell closure a study of \ce{^{200}Pt} is required, for which no \bdown value is known to date. Moreover, the Pt isotopes are only four valence protons below the $Z = 82$ proton shell closure, just below the Hg isotopes, which, when approaching the $N = 126$ shell, evolve towards sphericity with some degree of oblate deformation~\cite{Jolie01}. Hence, the \bdown values of Pt isotopes with respect to those of Hg isotopes are of high interest as both should approach the spherical limit towards $N = 126$, while only depending on the difference of two additional valence-proton holes, and may serve as a benchmark test of this neutron-shell closure. Furthermore, the $B(E2;4^+_1 \rightarrow 2^+_1)$ value test spherical versus deformed structures towards $N = 126$.

    In this manuscript, we report on the first measurement of the $B(E2;2^+_1 \rightarrow 0^+_1)$ and $B(E2;4^+_1 \rightarrow 2^+_1)$ values of \ce{^{200}Pt} via a direct measurement of the lifetimes of its first-excited $2^+$ and $4^+$ states.

 
    \section{Experiment}
	
    The \ce{^{198}Pt}(\ce{^{18}O}, \ce{^{16}O})\ce{^{200}Pt} two-neutron transfer reaction was used to populate low-lying excited states of \ce{^{200}Pt}. The \ce{^{18}O} beam of \SI{75}{\mega\electronvolt} was delivered by the \SI{9}{\mega\volt} tandem accelerator at the IFIN-HH in Bucharest-Măgurele and impinged on a \SI{600}{\micro\gram\per\centi\metre^2} thick self-supporting \ce{^{198}Pt} target. A gold foil was mounted parallel with respect to the target foil in order to stop the recoiling nuclei. Both foils were installed in the \nobreak ROSPHERE plunger device \cite{ref_rosphere} which is used to vary and stabilize the distance between target and stopper. $\gamma$\nobreakdash-ray spectroscopy data were taken at six different target-to-stopper distances, i.e., \SI{12}{\micro\metre}, \SI{40}{\micro\metre}, \SI{60}{\micro\metre}, \SI{80}{\micro\metre}, \SI{110}{\micro\metre} and \SI{150}{\micro\metre}; not including the plunger offset of \SI{-3.6 \pm 0.2}{\micro\metre} which was determined using the capacitance method \cite{ref_plungerbible, ref_capacitance_method}.
    
    
    The $\gamma$ radiation emitted from the decay of excited states was detected by 25 high-purity germanium (HPGe) detectors of the ROSPHERE array \cite{ref_rosphere} which were mounted in five rings. These rings were placed at angles of \SI{37}{\degree}, \SI{70}{\degree}, \SI{90}{\degree}, \SI{110}{\degree}, and \SI{143}{\degree} with respect to the beam axis. The back-scattered beam-like particles, i.e., \ce{^{16}O} and \ce{^{18}O}, were detected using the SORCERER particle detector array \cite{ref_sorcerer}. The data of the particle detectors allow the creation of particle-gated spectra which facilitate the separation of the occurring reaction channels and select the two-neutron transfer reaction. Due to the limited energy resolution of the solar cells and due to the similarity in kinematics, the two-neutron transfer reaction cannot be discriminated from the Coulomb excitation. However, the events of fusion evaporation are suppressed due to the different kinematics.
    
\begin{figure}[t]
		\includegraphics{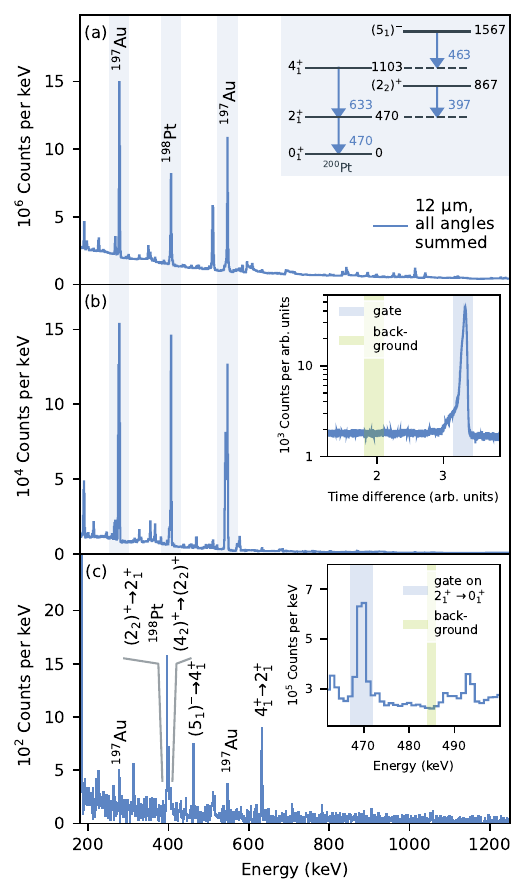}
		\caption{\label{raw_p_pg_spec} $\gamma$-ray energy spectra at \SI{12}{\micro\metre} with all angles summed up (a) without a coincidence condition, (b) with a coincidence condition set in the particle time-difference spectrum shown in the inset and (c) with both the particle gate from before and an additional coincidence condition set on the energy of the $2^+_1 \rightarrow 0^+_1$ transition shown in the inset. The suppression of fusion evaporation events by applying a particle gate can be seen from the difference between panels (a) and (b). The additional $\gamma$-ray energy gate on the \ce{{2}^+_1} $\rightarrow$ \ce{{0}^+_1} transition of \ce{^{200}Pt} reveals the feeding transitions but reduces the statistics too much to conduct a lifetime analysis. All observed transitions of \ce{^{200}Pt} are shown in the level scheme in panel (a).}
	\end{figure}

    Fig. \ref{raw_p_pg_spec} shows sum spectra of all detector rings taken at a target-to-stopper distance of \SI{12}{\micro\metre}, in particular the ungated $\gamma$-ray energy sum spectrum in panel (a), the particle-gated spectrum with the particle time-difference gate shown as an inset in panel (b), and a particle-$\gamma$ gated spectrum with a coincidence condition on the decay of the $2^+_1$~state of \ce{^{200}Pt}, the nucleus of interest, as shown in the corresponding inset in panel (c). From the difference between panels (a) and (b) the strong suppression of fusion-evaporation reactions is apparent, whereas the spectrum is dominated by $\gamma$-ray lines from Coulomb excitation, i.e., inelastic scattering. The additional $\gamma$-ray energy coincidence condition in panel (c) showcases the population of the isotope of interest, \ce{^{200}Pt}, however, with a modest level of statistics.

    The mean velocity of the recoiling \ce{^{200}Pt} ions after the target has been determined from the energies of the shifted and unshifted components of the $2^+_1 \rightarrow 0^+_1$ transition and amounts to $v/c = \SI{1.4 \pm 0.1}{\%}$.
    
    \section{Analysis}


    $\gamma$-ray emission from the $2^+_1$ state of \ce{^{200}Pt} was analyzed applying the recoil-distance Doppler-shift (RDDS) method and its lifetime was determined using the differential decay curve method (DDCM). Both methods are explained in detail in \linebreak Ref. \cite{ref_plungerbible}.
    
    In an RDDS experiment the nucleus of interest is produced in a nuclear reaction which populates excited states. The produced nuclei are accelerated out of the target foil in the direction of a stopper foil. The $\gamma$ decay of an excited state can either occur while the recoiling nuclei are in flight, which yields a Doppler-shifted peak in the energy spectra, or after they have come to rest in the stopper foil resulting in a peak at the $\gamma$-transition energy. The flight time between the target and the stopper foils is of the order of ten \si{\pico\second} and varies with the chosen target-to-stopper distance, serving as the time scale for the intended lifetime measurement and defining its sensitive range.
    
    The decay curve $R(t)$ is a time-dependent function of the intensities of the shifted $I_\text{s}(t)$ and unshifted $I_\text{u}(t)$ components of a $\gamma$-ray transition of interest,
    \begin{equation}
    \label{decaycurve}
        R(t) = I_\text{u}(t)/(I_\text{u}(t) + I_\text{s}(t)),
    \end{equation}
    where $t$ is the time of flight of the recoiling ions. $R(t)$ therefore depends on the time of flight $t$ of the nuclei of interest, given by their velocity $v$ and the set distance between the target and stopper foils. The lifetime of the $\gamma$-decaying excited state enters the shifted and unshifted intensities $I_\text{u,s}$. Data are normalized to the total number of counts in the prominent peak of the $5/2^+_1 \rightarrow 3/2^+_1$ transition of \ce{^{197}Au} at \SI{279}{\kilo\electronvolt} in order to compensate for beam fluctuations and varying measurement times.

    
    \begin{figure}[t]
		\includegraphics{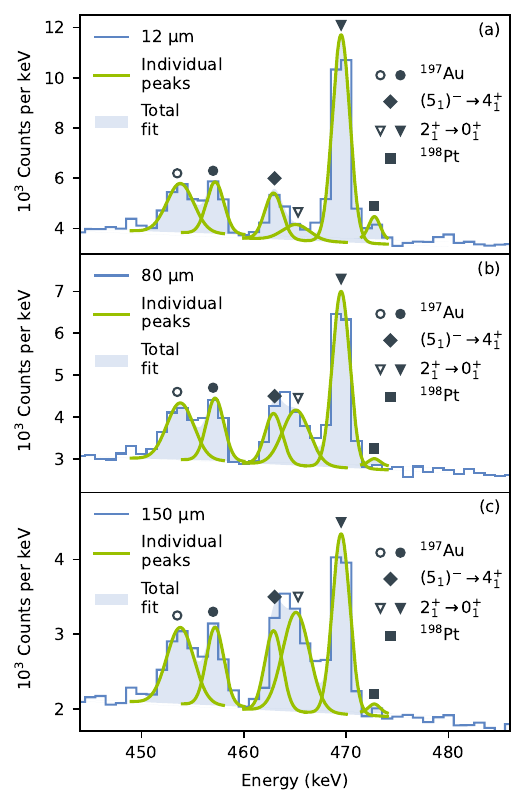}
		\caption{\label{dopplerspec} Particle-gated $\gamma$-ray singles spectra at \SI{143}{\degree} at \linebreak (a) \SI{12}{\micro\metre}, (b) \SI{80}{\micro\metre} and (c) \SI{150}{\micro\metre}. The origin of the shifted (\textit{open symbols}) and unshifted (\textit{full symbols}) components are indicated. The fits of the shifted and unshifted components of the \ce{{2}^+_1} $\rightarrow$ \ce{{0}^+_1} transition of \ce{^{200}Pt} and the relevant contaminants are shown. The evolution of the intensity ratio of the shifted and unshifted components with flight distance is visible, the intensity of the shifted component increases with the distance while the intensity of the unshifted component decreases.}
	\end{figure}

    The lifetime of the $2^+_1$ state can directly be obtained from the decay curve if the state is exclusively populated in the nuclear reaction and no feeding from higher-lying states is present. However, the $4^+_1$ and $(2_2)^+$ states of \ce{^{200}Pt} were populated in the 2n-transfer reaction, too, and decay into the $2^+_1$ state. In Fig. \ref{raw_p_pg_spec} (c) the feeding transitions $4^+_1 \rightarrow 2^+_1$ and \linebreak $(2_2)^+ \rightarrow 2^+_1$ are marked. These transitions become clearly visible after setting a coincidence condition on the $2^+_1 \rightarrow 0^+_1$ transition. The statistics of the particle-$\gamma$ gated data are not sufficient for a lifetime analysis using $\gamma$-$\gamma$ coincidences. Therefore, the analysis had to be performed using particle gated $\gamma$-ray singles spectra only. 

    The extraction of the shifted and unshifted intensities of the $2^+_1 \rightarrow 0^+_1$ transition of \ce{^{200}Pt} was hampered by the occurrence of other $\gamma$-ray transitions in the respective energy region. Figure \ref{dopplerspec} shows the fits of the shifted and unshifted components of the $2^+_1 \rightarrow 0^+_1$ transition as well as contaminating peaks originating from transitions of \ce{^{197}Au}, \ce{^{198}Pt} and \ce{^{200}Pt}. The former two are present due to Coulomb excitation of the target and stopper foils. For the observed $(5_1)^- \rightarrow 4^+_1$ transition of \ce{^{200}Pt} at \SI{463}{\kilo\electronvolt} no shifted component was observed even at large distances, hence, its lifetime is larger than the sensitivity of the experiment. Note that the lifetime of the corresponding state of \ce{^{196}Pt} is known to be $\tau(5^-_1, \ce{^{196}Pt}) = \SI{1.1 \pm 0.2}{\nano\second}$. Therefore, only an unshifted 463-keV transition needed to be considered for the fit of the present data. The intensity of the $(3)^+_1 \rightarrow 2^+_2$ transition of \ce{^{198}Pt} at \SI{473}{\kilo\electronvolt} was found to be sufficiently small at all distances and angles to neglect its potential shifted component in the analysis. The intensities of the shifted and unshifted components of the $7/2_2^+ \rightarrow 5/2_1^+$ transition of \ce{^{197}Au} at \SI{458}{\kilo\electronvolt} were disentangled from the higher-lying transitions in the energy region of interest via consistent fits of the data at all angles and distances. Through this iterative procedure, shifted and unshifted intensities of the $2^+_1 \rightarrow 0^+_1$ transition of \ce{^{200}Pt} were extracted.
    
    The correction for the above-mentioned slow feeding from the $4^+_1$ and $(2_2)^+$ states observed in the particle-$\gamma$ gated spectra was done by subtraction \cite{Kocheva_2024}. The intensities of the unshifted component of the feeding transitions were subtracted from the intensities of the unshifted part of the $2^+_1 \rightarrow 0^+_1$ transition of \ce{^{200}Pt}.

\begin{figure}[t]
		\includegraphics{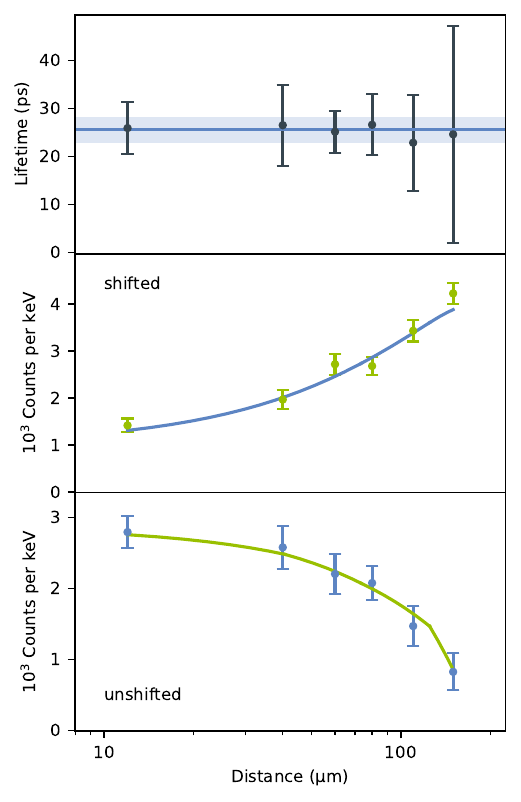}
		\caption{\label{napa} Exemplary fits obtained in the DDCM analysis for the \SI{143}{\degree} detector ring. The \textit{top} panel shows the resulting lifetime values for each distance setting and their weighted mean value and its uncertainty indicated by a horizontal line and the shaded area, respectively. The \textit{centre} and \textit{bottom} panel illustrate the normalized intensity of the shifted and unshifted components with the resulting second degree spline function and derivative of the spline function, respectively.}
	\end{figure}

    The DDCM analysis for the $2^+_1$ state of \ce{^{200}Pt} was performed using the contaminant-free and feeding-corrected $\gamma$\nobreakdash-ray intensities. Its lifetime $\tau$ is determined for each distance following 
    \begin{equation} \label{ddcmeq}	
		\tau = \frac{I_\text{u}}{\frac{\text{d}}{\text{d}t} I_\text{s}} = \frac{I_\text{u}}{v \, \frac{\text{d}}{\text{d}x} I_\text{s}}.
	\end{equation}
    The evolution of the intensities of both components with the distance is shown in Fig. \ref{dopplerspec}. Due to the differential approach of the DDCM only relative distances are required. The mean lifetime $\tau$ is determined via a $\chi^2$~fit of the distance-wise determined lifetime values implemented in the program \texttt{napatau} \cite{ref_napatau}, separately for each ring of ROSPHERE. An exemplary fit which entered the determination of the final value of $\tau$ is shown in Fig. \ref{napa}. Since the shifted and unshifted components of the transition are related through Eq. (\ref{ddcmeq}) via the lifetime as a free parameter, both observables are fitted simultaneously by a quadratic spline function in \texttt{napatau}. The weighted average of all distance values included in the fit is then determined for each detector ring.

\begin{figure}[t]
	\includegraphics{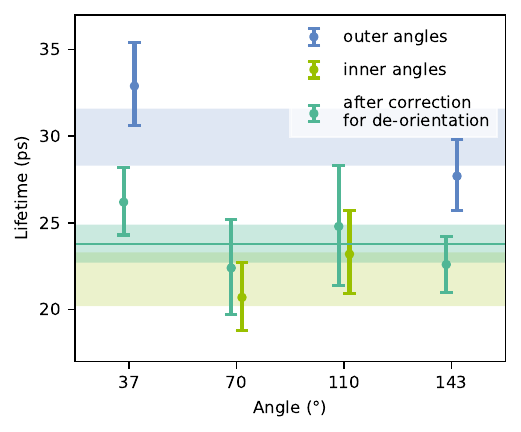}
	\caption{\label{lifetimes_deo} Lifetime values before and after correcting the de-orientation effect. The data points representing the outer detector angles of \SI{37}{\degree} and \SI{143}{\degree} (\textit{blue}) as well as the inner angles of \SI{70}{\degree} and \SI{110}{\degree} (\textit{green}) agree with their respective mean values but not overall. The lifetime values after the thus necessary de-orientation correction of each detector angle along with their mean value are shown in \textit{cyan}.}
	\end{figure}

    The thus obtained lifetime values for each detector ring angle are included in Fig. \ref{lifetimes_deo}. In particular when averaging the data from each corresponding pair of rings relative to \SI{90}{\degree}, \SI{37}{\degree}/\SI{143}{\degree} (\textit{blue}) and \SI{70}{\degree}/\SI{110}{\degree} (\textit{green}), it is apparent that the lifetime values for the respective inner and outer angles do not match, as indicated by the respective weighted averages for both angle pairs, and vary by almost \SI{10}{\pico\second}, with a total average of \SI{26 \pm 2}{\pico\second}.

\begin{figure}[t]
		\includegraphics{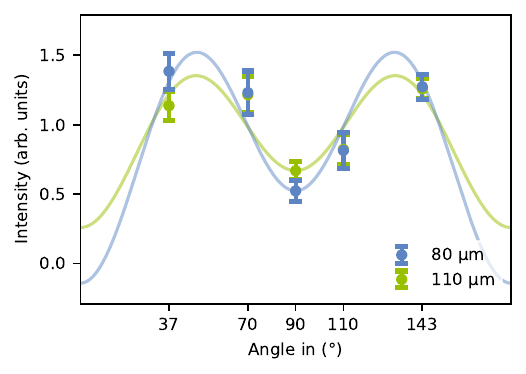}
		\caption{\label{deor_attenuation} Angular distributions of the $2^+_1 \rightarrow 0^+_1$ transition at distance settings of \SI{80}{\micro\metre} (\textit{blue}) and \SI{110}{\micro\metre} (\textit{green}), each normalized and fitted by the angular distribution function according to Eq. \eqref{eq:angdist}.
        }
	\end{figure}

    A potential source of this disagreement can be time-dependent angular distributions, which can be attributed to the de-orientation effect. This effect has frequently been used for the measurement of $g$ factors using recoil-into-vacuum techniques \cite{Stuchbery07, Radeck12}. After the nuclear reaction, which produces \ce{^{200}Pt} in its excited states, the spins of the \ce{^{200}Pt} nuclei are aligned to some extent. The alignment leads to a distinct angular distribution of the decay-$\gamma$ rays. Due to the hyperfine interaction of the oriented nuclear spin and the random electronic spin of the remaining electrons of the recoiling ions, leading to a precession of the nuclear spins about the (randomly oriented) total spin, the initial angular distribution is attenuated with time. An analysis of the total angular distributions of the $2^+_1 \rightarrow 0^+_1$ transition at the various distances, despite little available statistics, shows a dependence of the angular distribution on time (i.e., distance), as shown for two exemplary settings in Fig. \ref{deor_attenuation}. The angular distributions were obtained by fitting the sum intensities with the normalized angular-distribution function
    \begin{equation}
    \label{eq:angdist}
        \frac{W(\vartheta)}{A_0} = 1 + \sum_{i=2,4} \frac{A_i}{A_0} P_i(\cos\vartheta),
    \end{equation}
    with the Legendre polynomials $P_i$, normalization parameter $A_0$ and fit parameters $A_i$.
   
    An effective way to correct the lifetime measurement for the de-orientation effect is to apply correction factors to the intensities, which renormalize the angular distributions at each distance setting to an isotropic distribution. Since the nuclei which de-excite in the stopper are longer exposed to the hyperfine interaction, de-orientation affects the unshifted component by more than an order of magnitude stronger than the shifted one. Therefore, the correction factors are only applied to the intensities of the unshifted component of the $2^+_1 \rightarrow 0^+_1$ transition of \ce{^{200}Pt}. The resulting lifetimes for all rings and their weighted average are included in \textit{cyan} in Fig. \ref{lifetimes_deo}. After the de-orientation correction the data of all rings match within their uncertainty, and the weighted average of $\tau(2^+_1,\ce{^{200}Pt}) = \SI{23.8 \pm 1.1}{\pico\second}$ is adopted as the final result.
    Using a conversion coefficient of $\alpha = \num{0.02892 \pm 0.00002}$ \cite{bricc} the corresponding $E2$ transition strength results in
    \begin{equation*}
    	B(E2; 2^+_1 \rightarrow 0^+_1) = \asymUnc{1460}{80}{70}{e^2\femto\metre^4} = \asymUnc{21.0}{1.1}{1.0}{\weisskopfunit}
    \end{equation*}

    As the statistics for the $4^+_1 \rightarrow 2^+_1$ transition is limited, the lifetime of the $4^+_1$ state is determined as described in Ref. \cite{litzinger_2015}. The ratio of the unshifted and shifted components fitted in the summed spectra of all distance settings $R_\text{sum}$ is used to calculate the lifetime of the $4^+_1$ state using
    \begin{equation}
        R_\text{sum} = \frac{I_u}{I_u + I_s} = \sum_i n_i R(x_i)
    \end{equation}
    with the intensities of the shifted and unshifted components of the sum spectrum $I_s$ and $I_u$, the normalization factors $n_i$ for each distance $i$ and the decay curve $R$ which can be expressed as the solution of the Bateman equations \cite{Bateman10}. Feeding of the $(5_1)^- \rightarrow 4^+_1$ transition is taken into account as described before. Further feeding transitions are not visible in the spectra but are considered as \SI{10}{\%} unobserved feeding as a systematic uncertainty.
    This procedure results in a lifetime of $\tau(4^+_1,^{200}$Pt$) = \SI{2.6 \pm 0.3}{\pico\second}$ and a corresponding quadrupole transition strength of
    \begin{equation*}
    	B(E2; 4^+_1 \rightarrow 2^+_1) = \asymUnc{3040}{430}{340}{e^2\femto\metre^4} = \asymUnc{44}{7}{5}{\weisskopfunit}
    \end{equation*}
    The lifetimes measured in this work and their resulting $B(E2)$ values are summarized in Table \ref{tab_results}.

    \section{Discussion}
    
    As the Pt isotopic chain is approaching the $N~=~126$ major shell closure, a transition from quadrupole-deformed, $\gamma$-soft structures as present in, e.g., \ce{^{196}Pt} to spherical shapes is expected. A well-known indicator for nuclear structure is given by the energy ratio
    \begin{equation}
    \label{eq:r42}
        R_{4/2} = \frac{E(4^+_1)}{E(2^+_1)}
    \end{equation}
    with benchmark values of $R_{4/2} = 3.33$ for well-deformed rotors, $R_{4/2} \approx 2.5$ for the $\gamma$-soft limit, and $R_{4/2} = 2$ in the spherical vibrational limit. For \ce{^{200}Pt} one obtains $R_{4/2}(^{200}$Pt$) = 2.34$, which is located right between these benchmark values, hence, on a transition between deformed $\gamma$-soft to spherical shapes. For the low-lying levels of a near-spherical, vibrational nucleus, one expects a near-degenerate triplet of $4^+_1$, $2^+_2$, and $0^+_2$ states. A tentative $(0^+_2)$ state at \SI{1118}{\kilo\electronvolt}, very close to the $4^+_1$-state energy of \SI{1103}{\kilo\electronvolt}, has been observed in only one experiment \cite{Yates88}. The known $(2_2)^+$ state at an energy of \SI{868}{\kilo\electronvolt}, however, is a bit lower than the other two potential members of the triplet. Therefore, level energies are not quite conclusive on the low-lying structure of \ce{^{200}Pt}.

    \begin{table}[t]
		\caption{\label{tab_results} Resulting lifetimes and reduced transition strengths for the \ce{{4}^+_1} $\rightarrow$ \ce{{2}^+_1} and \ce{{2}^+_1} $\rightarrow$ \ce{{0}^+_1} transitions of \ce{^{200}Pt} determined in this work.}
        \vspace{5pt}
	\begin{ruledtabular}
		\begin{tabular}{lccc}
            \multirow[t]{2}{*}{Transition $J_i \rightarrow J_f$} & \multirow[t]{2}{*}{Lifetime} & \multicolumn{2}{c}{$B(E2; J_i \rightarrow J_f)$}\\
            \noalign{\vskip 2pt}
            & in \si{\pico\second} & in \si{e^2\femto\metre^4} & in \si{\weisskopfunit} \\ \hline
            \noalign{\vskip 2pt}
            $2^+_1 \rightarrow 0^+_1$ & \num{23.8 \pm 1.1} & \asymUnc{1460}{80}{70}{} & \asymUnc{21.0}{1.1}{1.0}{} \\ \noalign{\vskip 2pt}
            $4^+_1 \rightarrow 2^+_1$ & \num{2.6 \pm 0.3} & \asymUnc{3040}{430}{340}{} & \asymUnc{44}{7}{5}{} \\
        \end{tabular}
	\end{ruledtabular}
	\end{table}

    A more direct probe of the wave functions of the respective states is offered through the now obtained $B(E2)$ values, and the ratio 
    \begin{equation}
    \label{eq:b42}
        B_{4/2} = \frac{B(E2; 4^+_1 \rightarrow 2^+_1)}{B(E2; 2^+_1 \rightarrow 0^+_1)}
    \end{equation}
    takes typical values of $B_{4/2} = 1.4$ in the axially-symmetric and $\gamma$-soft deformed limits, and $B_{4/2} = 2$ in the spherical limit. From the values given in Table \ref{tab_results} one obtains $B_{4/2} = \num{2.08 \pm 0.32}$, which is in very good agreement with the spherical limit.

\begin{figure}[t]
		\includegraphics[trim = 5pt 0 0 0]{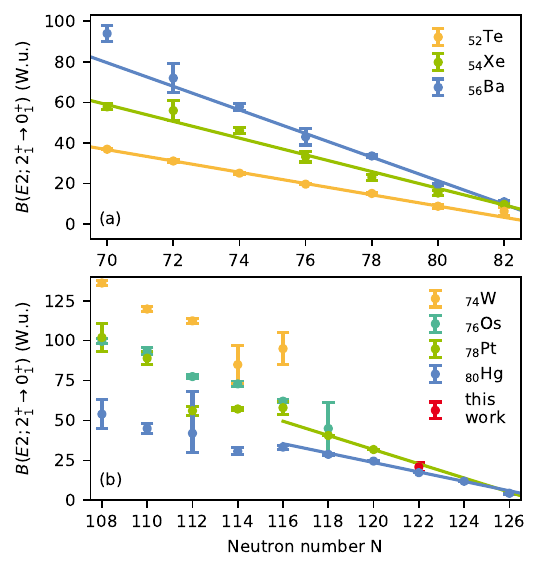}
		\caption{\label{be2} Reduced transition strengths \bdown for the (a) Te, Xe and Ba and (b) W, Os, Pt and Hg isotopic chains along with linear fits for the trend of the \bdown value for Te, Xe, Ba, Pt and Hg. The proximity of the two respective regions of the nuclear chart to the shell closures at $Z = 50$ and $N = 82$ as well as $Z = 82$ and $N = 126$ allows investigating mirror effects. The linear fits can be used to determine the effective charges within the U(5) limit of the IBM using Eq. \eqref{eq_effcharge}. The data is taken from \cite{nds_122, nds_124, nds_126, nds_128, nds_130, nds_132, nds_134, nds_136, nds_138, 126-128xe_be2, 130xe_be2, nds_182, nds_184, nds_186, nds_188,nds_190, nds_192, nds_194, nds_196, nds_198, nds_200, nds_202, nds_204, nds_206, Sahin24, 192hg_be2}.}
	\end{figure}

    To further elucidate the evolution of collective structures in the Pt isotopic chain, we compare its \bdown values to those of neighbouring isotopic chains in Fig.~\ref{be2}~(b). All displayed isotopic chains show \bdown values dropping towards $N = 126$, while there is a flattening of \bdown values around $N = 114, 116$, in particular for the Pt isotopes. Tentatively, such a trend is also observed in the Hg and W isotopic chains and had been discussed in view of the prolate-oblate shape transition and subsequent approach to the $N=126$ shell closure~\cite{Sahin24}. Past $N = 116$, $B(E2)$$\downarrow$~values drop near-linearly.

    A similar trend is known from other regions of the nuclear chart, in particular isotopes near the $Z=50$ proton shell, i.e., Te, Xe and Ba isotopes towards $N=82$, which can be viewed as a valence mirror region \cite{ValenceMirror} to the Os, Pt and Hg isotopes, replacing proton particles with proton holes. The linear trend of the \bdown values, shown in Fig. \ref{be2} (a), which is pronounced in the Te, Xe, and Ba isotopes, can be attributed to the spherical-vibrational character of these nuclei. In the vibrational limit, as given by the U(5) limit of the IBM, the \bdown value is given by the simple relation
    \begin{equation}
    \label{eq_effcharge}
        B(E2; 2^+_1 \rightarrow 0^+_1) = e_\text{B}^2 \ N_\text{B}
    \end{equation}
    with the effective boson number $N_\text{B}$ and the effective charge $e_\text{B}$, which can be determined from the slope of the data. The results of linear fits to the \bdown values along the Te, Xe, Ba, Hg and Pt isotopic chains in Table~\ref{tab_eff_charges} show similar values for isotopes with the same amount of valence protons and holes. A difference between both nuclear regions, however, is that \bdown values converge towards the neutron shell closure at $N = 82$ in the lighter isotopes, whereas in the heavier region \bdown values are already very close to each other at $N = 122$, four neutrons below the next neutron shell closure. This behaviour may be attributed to the specific underlying shell structure of the isotopes towards \ce{^{208}Pb}.

    \parfillskip=0pt

    We note that the top orbitals of the proton shell, which are filled towards $Z=82$, are the $\pi 2s_{1/2}$ and the $\pi 2d_{3/2}$ orbitals, which can be inferred from the lowest-lying states of heavier odd-proton isotopes. The low spin of these orbitals leads to a suppression of collectivity, in particular, the $s_{1/2}$ orbital cannot significantly contribute to collectivity in the $2^+$ or $4^+$ wave functions. In the neutron sector, the dominant orbitals below $N=126$ are the $\nu 3p_{1/2}$ and the $\nu 2f_{5/2}$ orbitals. Again, a $p_{1/2}$ orbital hinders collectivity towards the major shell closure.

    \parfillskip=0pt plus 1fil
    
    \begin{table}[H]
	\caption{\label{tab_eff_charges} Effective charges determined using a linear fit to the trends of the \bdown value shown in Fig. \ref{be2}. Note the similarity of the results for nuclei with the same amount of valence protons (or holes).}
    \vspace{5pt}
	\begin{ruledtabular}
        \begin{tabular}{ccc}
    \vspace{5pt}
            Isotope & $e_\text{B}$ in $\sqrt{\si{\weisskopfunit}}$ & \# valence protons (holes)\\ \hline
            \noalign{\vskip 2pt}
            \ce{_{52}Te} & \num{1.67 \pm 0.01} & 2\\
            \ce{_{54}Xe} & \num{2.03 \pm 0.01} & 4\\
            \ce{_{56}Ba} & \num{2.41 \pm 0.04} & 6\\
            \noalign{\vskip 2pt}
            \hline
            \noalign{\vskip 2pt}
            \ce{_{80}Hg} & \num{1.72 \pm 0.01} & 2\\
            \ce{_{78}Pt} & \num{2.11 \pm 0.02} & 4\\
        \end{tabular}
	\end{ruledtabular}
    \end{table}
    \noindent Therefore, the effective valence space for the formation of the low-lying $2^+$ and $4^+$ states does not simply scale with the number of valence nucleons, but is suppressed by about one pair of active valence proton and neutron holes, each. Consequently, the occurrence of $j=1/2$ orbitals could be responsible for the rather small difference in collectivity between Hg and Pt isotopes with 2 and 4 valence-proton holes, respectively, and 4 valence-neutron holes, each.
    
    
    \section{Conclusion}

    The lifetimes of the $2^+_1$ and $4^+_1$ states of \ce{^{200}Pt} were measured for the first time.  The recoil-distance Doppler-shift (RDDS) method was applied. \ce{^{200}Pt} was produced in a two-neutron transfer reaction at the tandem accelerator at the IFIN-HH. The $B(E2)$ values for the $2^+_1 \rightarrow 0^+_1$ and $4^+_1 \rightarrow 2^+_1$ transitions were determined enabling the determination of the $B_{4/2}$ ratio. The obtained $B_{4/2}$ ratio is in good agreement with the theoretical limit of a vibrational nucleus indicating the structural evolution towards sphericity as the Pt isotopic chain approaches the neutron-shell closure at $N = 126$. 
    
    \begin{acknowledgments}
    We are indebted to the staff of the IFIN-HH tandem accelerator for providing optimal beam conditions. We further thank Arwin Esmaylzadeh for valuable discussions on the analysis, Georgi Rainovski for fruitful discussions on the structural evolution in this region, and Andrew Stuchbery for an exchange on nuclear de-orientation. This work was supported by the Deutsche Forschungsgemeinschaft (DFG, German Research Foundation) as part of the Project\mbox{-}ID~264883531 – Research Training Group 2128 'Accelence' and Project\mbox{-}ID~499256822 – Research Training Group 2891 'Nuclear~Photonics' and by the German Federal Ministry of Education and Research (BMBF) under Grant Nos. 05P19RDFN1, 05P21RDFN1, 05P21RDFN9, and 05P24RD3. This project has received funding from the European Union’s Horizon 2020 research and innovation programme under grant agreement No. 654002 (ENSAR-2).
    
    \end{acknowledgments}
    
    \appendix

    \nocite{*}
    
    \bibliography{200Pt_paper}
	
\end{document}